%% file: icassp4_261003.tex
\documentclass[9pt]{article}
\usepackage[utf8]{inputenc}
\usepackage{spconf,amsmath,amssymb,graphicx,hyperref}
\usepackage{cite} % sorted, compressed citation ranges, e.g. [3--6]; leaves spconf's reference list untouched
\usepackage[protrusion=true,expansion=true]{microtype}
\usepackage{booktabs}
\usepackage{multirow}
\usepackage{array}
\usepackage[table]{xcolor}
\definecolor{palTeal}{HTML}{31707A}
\definecolor{palSky}{HTML}{9FE0F6}
\definecolor{palYellow}{HTML}{F3E59A}
\definecolor{palPeach}{HTML}{F3B59B}
\usepackage{booktabs}
\usepackage{tabularx}
\usepackage{threeparttable}

\usepackage{tikz}
\usetikzlibrary{shapes.geometric,arrows.meta,positioning,calc,fit,backgrounds,decorations.pathreplacing}

\title{Can LLM Agents Automate Reinforcement Learning for Text-to-Speech?}

\name{Xuanjun Chen$^{1, 2*}$\thanks{* Work done during internship at Shanda Group.}, Zixiong Su$^{2}$, Hao Shi$^{2}$, Chang Zeng$^3$, Kai Li$^4$, Jyh-Shing Roger Jang$^{1}$, Hung-yi Lee$^{1}$}
\address{$^1$National Taiwan University,
$^2$Shanda Group, $^3$National Institute of Informatics, 
$^4$Tsinghua University %\quad 
}

\begin{document}
\ninept
\maketitle

\begin{abstract}
Although reinforcement learning (RL) post-training repairs the localized segmental errors of zero-shot text-to-speech (TTS), arriving at a working recipe still relies on tedious manual tuning, and whether LLM agents can take over this research pipeline is unclear. 
We investigate this question with AgenticTTS-Forge, a collaborative workflow that structures human guidance and agentic execution around a shared workspace, applied to CosyVoice2-0.5B. 
To measure what the agent automates, we audit its trajectory stage by stage against the published recipe. To measure what it exploits, we score its policies with held-out observers hidden from the agent. Our results show that the agent recovers an underspecified recipe, improves it, and, when gains stall, surveys the literature unprompted and pivots from the LM carrier to the flow carrier, halving Bad cases. However, its autonomy exposes three traps across the data, proxy, and algorithm axes: the held-out set leaks through a channel the contract never reads, a self-shaped reward inflates the proxy where it is scored, and separately tuned policies do not compose additively. These findings show that the binding constraint is measurement rather than reasoning, and can inform the design of harnesses whose contracts read every channel the agent does.
\end{abstract}


\section{Introduction}

While large language models (LLMs) have demonstrated the critical role of post-training in systematically aligning model behavior with human preferences and driving major gains in core capabilities, the Text-to-Speech (TTS) domain faces a parallel necessity. 
Recent advancements in neural codec-based TTS have enabled impressive zero-shot speech synthesis by scaling up model parameters and training data~\cite{anastassiou2024seedtts,du2024cosyvoice2}. However, while this scaling has largely mitigated systemic errors like poor speaker similarity, modern TTS systems continue to suffer from fine-grained and localized quality degradations. These issues, referred to as segmental errors, include mispronunciations, abnormal pauses, and semantic-phonetic alignment problems~\cite{fpo2025}.

Because these localized issues often occur in specific segments of otherwise well-generated audio, they are not effectively addressed by standard global training objectives. Consequently, post-training emerges as an essential paradigm for TTS development. 
Existing supervised fine-tuning (SFT) methods enable targeted re-synthesis~\cite{liu2026emotra,song2026diagnose}, but they are constrained by their reliance on explicit target audio and fixed data distributions. 
By integrating reinforcement learning (RL) and human preference data to calibrate pre-trained models~\cite{fpo2025,zhang2024speechalign,chen2024uno,hu2024rio,yu2026listen}, this RL paradigm shifts the optimization focus to targeted local refinement. This alignment process is crucial for correcting problematic regions, significantly reducing the bad case rate, and ultimately ensuring the robustness and perceptual quality of zero-shot TTS systems. 

Autonomous AI agents can now execute complex workflows, write code, and operate developer tools with minimal human oversight~\cite{lu2024aiscientist,huang2024mlagentbench,chan2024mlebench}, offering a powerful way to automate research and development. RL Post-training is specifically well suited for this automation because it is a clearly defined stage that produces direct and measurable evaluation signals. By moving away from manual trial and error, these agents utilize their broad autonomy to independently curate data, build training pipelines from scratch, and iteratively refine optimization strategies. Evidence shows that on focused tasks with clear metrics, autonomous agents can carry post-training end to end, though still short of expert engineering~\cite{posttrainbench2026,ftdojo2026}. Integrating this agentic workflow into TTS systems provides a highly scalable method to automatically correct localized audio errors and improve overall performance.  

\input{Figure/figure1new}

Whether such agents can automate the full research pipeline that human engineers follow, reproducing a baseline, improving it, and pivoting architectures, remains unclear for TTS. Therefore, this paper aims to investigate the overarching question: \textit{``Can LLM agents automate reinforcement learning for text-to-speech?''} 
As illustrated in Fig.~\ref{fig:system}, we introduce AgenticTTS-Forge, a human-agent collaborative workflow in which humans provide high-level guidance while the LLM agent executes the human-like research process. 
To answer this question, our work provides three main contributions:

\begin{itemize}
\item We run the end-to-end RL post-training pipeline to demonstrate that an LLM agent can discover valid RL recipes and, when optimization plateaus, survey the literature unprompted to shift its optimization focus from the LM Carrier (the token LM) to the Flow Carrier (the flow-matching decoder). 
\item Through auditing the agent's trajectory, we identify three traps of autonomy, one per axis of the RL design space: a held-out set leaking through an unread channel (data), a self-shaped reward inflating its proxy (proxy), and separately tuned policies whose gains do not add up (algorithm).
\item We offer novel insights into agentic research by auditing the shared workspace. We find that the binding constraint is measurement rather than reasoning: preventative rules held only where the contract reached, and post-hoc auditability recovered what prevention missed.
\end{itemize}

\section{The TTS Model Preliminaries for RL}
\label{sec:preliminaries}
Our RL post-training is built upon CosyVoice2~\cite{du2024cosyvoice2}. 
During RL optimization process, existing utterances act as reference prompts to generate new speech, providing the necessary training signals.

\textbf{Base TTS Architecture. } CosyVoice2 decouples speech synthesis into semantic and acoustic stages through four modules. The process begins with a speech tokenizer, followed by a unified text-speech Language Model (LM) responsible for predicting semantic tokens. A chunk-aware Flow Matching (FM) model then renders them into a Mel spectrogram under the reference conditions, and a vocoder converts it to a waveform. We freeze the tokenizer and vocoder and optimize only the text-speech LM and the FM module.

\textbf{RL Design Space for TTS. } We decouple the RL pipeline into three axes: data, proxy, and algorithm. The \emph{data} axis fixes which conditions and generated candidates enter training and how they are arranged for the update (e.g., preference pairs vs. sampled groups). The \emph{proxy} provides the scalar reward signal, utilizing either trained human-preference models or composite objective metrics (e.g., Automatic Speech Recognition (ASR), Mean Opinion Score (MOS)). Finally, the \emph{algorithm} dictates how the policy is updated from that signal: pairwise Direct Preference Optimization (DPO)-style losses~\cite{rafailov2023dpo} on the LM, or critic-free Group Relative Policy Optimization (GRPO)~\cite{shao2024deepseekmath} with intra-group relative advantages on the decoder. We refer to these two optimization targets as the LM Carrier and the Flow Carrier. 

\section{The Proposed Workflow and Details}
\label{sec:method}
Fig.~\ref{fig:system} illustrates the AgenticTTS-Forge, featuring two layers: a four-stage workflow (top) where human and agent collaborate, and an shared workspace (bottom) that supports continuous interaction. 

\textbf{Human-Agent Collaboration.} The workflow relies on an iterative interaction between human guidance and agent execution across four stages, named here by the human's prompting role and in Sec.~\ref{sec:autonomy} by the agent's action (Table~\ref{tab:master_trajectory_summary}). In Stage 1 (Baseline Specification), the human anchors the task via reference literature, setting the initial target for the agent to reproduce. In Stage 2 (Trajectory Guidance), human prompting effort peaks to steer step-wise optimization, guiding the agent through design choices left unspecified by the paper. When optimization plateaus, the human role diminishes in Stage 3 (Pivot Authorization): rather than prescribing methods, the human merely flags the symptom and delegates the solution, authorizing the agent's proposed methodological shift. 
% Finally, in Stage 4 (Integration Approval), the human selects the final composition from the agent's scale sweep. 
Finally, in Stage 4 (Integration Approval), the human explicitly selects the optimal final composition derived directly from the agent's scale sweep. 

\input{Table/table_roles}
\input{Table/table_main}
\textbf{The Shared Workspace.} To guide decision-making throughout these stages, the shared workspace relies on two main components: a measurement contract that sets strict boundaries, and a trajectory memory that logs past choices to guide subsequent iterations.

\vspace{1mm}
\noindent\textit{(a) Measurement Contract (Extensible Rules).}
The contract predefines both the final evaluation benchmarks and the initial in-loop proxy models used during RL. By decoupling the evaluation metrics from the training proxies, the contract acts as a gatekeeper to prevent the agent from exploiting the metrics; it gated every LM candidate (the LM pairs / gate column of Table~\ref{tab:roles}), whereas the proxies the agent later introduced at its pivot never passed through it (Sec.~\ref{sec:proxy_exploitation}). It defines three rule categories, instantiated in Table~\ref{tab:roles}:
\begin{itemize}
    \item \textit{Constraints (Prerequisites)}: These enforce basic validity through \emph{Isolation} and \emph{Floors}. \emph{Isolation} pins variables such as tool versions and random seeds so they cannot confound the analysis. \emph{Floors} reject candidates that circumvent task requirements (e.g., bypassing audio synthesis) or fail to outperform a seed-varied baseline in both languages.
    \item \textit{Proxies (In-loop)}: Corresponding to the In-loop block of Table~\ref{tab:roles}, these models serve as training proxies. The LM carrier (Stage 1) requires no reward model, relying directly on preference pairs (LM pairs) for its FPO loss, with all candidates subject to a 512-condition gate (LM gate). Later, the flow carrier, adopted during the Stage 3 pivot, optimizes a weighted sum of its three reward entries (Flow reward).
    \item \textit{Observers (Held-out)}: As detailed in the held-out block of Table~\ref{tab:roles}, these are held-out evaluation models that assess the TTS performance using a sealed test set intentionally withheld by design (Sec.~\ref{sec:traps} discusses instances where this failed). Any partial overlaps are indicated in the table footnotes.
\end{itemize}

\vspace{1mm}
\noindent\textit{(b) Trajectory Memory (Runtime Log).}
Rather than lossy textual summaries~\cite{lee2026metaharness}, every evaluated candidate is appended to a runtime log holding its \textit{Recipe} (executable code), \textit{Score}, and \textit{Trace}, which records \emph{why} a candidate failed, not merely that it did. Stable indices ($i_k$ LM candidates, $o_k$ flow candidates, $g_k$ retrieved groundings) keep the trajectory auditable (Table~\ref{tab:master_trajectory_summary}) and let the Stage 3 pivot be derived from the record rather than prompted (Sec.~\ref{sec:method_search}). 

\section{Experiments}
\label{sec:exp}

All systems initialize from CosyVoice2-0.5B~\footnote{\href{https://huggingface.co/FunAudioLLM/CosyVoice2-0.5B}{https://huggingface.co/FunAudioLLM/CosyVoice2-0.5B}} and are evaluated on Seed-TTS-Eval~\cite{anastassiou2024seedtts}~\footnote{\href{https://github.com/BytedanceSpeech/seed-tts-eval}{https://github.com/BytedanceSpeech/seed-tts-eval}} (test-zh 2020, test-en 1088, test-hard 400) by word and character error rate (WER, CER), speaker embedding cosine similarity (SECS), and MOS. Both carriers search a shared, hash-pinned pool of 5,000 conditions (AISHELL-3, LibriTTS) using FPO's 100 prompt speakers~\cite{fpo2025} with 8 samples/condition. Training and evaluation are separated at the corpus level.

\subsection{Agent Outcomes: What Was Automated, and at What Cost}
\label{sec:overall_performance}
\textbf{The pipeline operated end to end: every gain in the lower block of Table~\ref{tab:main} came from the agent's own exploration, with no training recipe edited by hand; the one hand-set value is the merge scale (Sec.~\ref{sec:autonomy}).} Read stage by stage, the trajectory shows the agent breaking down an optimization problem that expert recipes solve with a single LM carrier policy. Stage~1 ($i_{10}$) recovers the direction of the published FPO~\cite{fpo2025} effect on every intelligibility set. The Stage~2 LM carrier policy ($i_{20}$) then consistently improves intelligibility across all test sets while leaving speaker similarity unchanged. 
The Stage~3 flow carrier policy ($o_{23}$) targets a complementary objective: it raises UTMOSv2 naturalness on every set and halves the bad case rate, with intelligibility held at the base. 
Composition (Stage~4) merges the two policies without training and retains the highest naturalness and the lowest bad case rate of the block. Measured by the held-out observers, then, the pipeline succeeded; the in-loop dynamics behind these numbers are less clean, and they set the agenda for the rest of this section. 
Sec.~\ref{sec:autonomy} reconstructs from Table~\ref{tab:master_trajectory_summary} how an agent given only the FPO paper arrived at $i_{20}$ and $o_{23}$. Sec.~\ref{sec:traps} examines what Table~\ref{tab:main} alone cannot show: a test set that leaked through a channel the contract never read, a reward the agent inflated where it was scored, and a merge that gained naturalness but sacrificed speaker similarity and part of $i_{20}$'s intelligibility gain. Sec.~\ref{sec:control} asks which workspace mechanisms contained these failures and which did not.

\input{Table/table1_condensed_v3}
\input{Table/table_contrast_v2}

% \subsection{Workflow Dynamics: What the Agent Solved on Its Own}
\subsection{Workflow Dynamics: What the Agent Achieved on Its Own}
\label{sec:autonomy}\label{sec:recipe_diff}\label{sec:data_dynamics}\label{sec:method_search}
Table~\ref{tab:master_trajectory_summary} records the four stages of the workflow. Read against its Published Recipe column, the agent first constructed the structural space, then executed a heuristic search that revised the recipe, redefined the search space by transferring an external method, and finally swept integration scales for a post-hoc composition. Every exploratory move was the agent's; only the final scale was ours.

\textbf{Stage 1: Construct Structural Space. } FPO leaves most settings unstated. With the paper as the only method prompt, the agent rewrote the recipe as a family around FPO, each unstated choice a search coordinate, and settled the data side by measurement: a pilot, an extrapolated pair budget, and its own gates. The result matched the published effect in direction, not size.

\textbf{Stage 2: Execute Heuristic Search. } Under our heaviest steering, one falsifiable mutation per round, the agent searched that family along the algorithm and data axes. Reading its own loss curves, it diagnosed unequal pair lengths as the confounder and revised the objective at two coordinates; later objective edits were rejected by the contract's 512-condition gate. The gain came from the data axis: pairs generated once from the initial policy had gone stale, so the agent regenerated them on-policy, and that cut, not the objective edits, improved intelligibility over the base.

\textbf{Stage 3: Redefine Search Space. } When the LM carrier ran out of gains, the agent pivoted to the flow carrier under the least prompting of any stage: we named a symptom, a Bad case tail that no LM edit moved, and no method. It located the bottleneck in the decoder, surveyed the literature unprompted, and transferred FlowTTS-GRPO~\cite{flowttsgrpo2026} from CosyVoice3 to CosyVoice2, a redefinition, not a copy: a stochastic ordinary differential equation (ODE) sampler so that GRPO has a log-probability to score, a different naturalness proxy, and conditions banded by the base's own scores. The pivot revived nearly every dead condition and halved Bad cases; the published configuration, which we could complete only on CosyVoice3, lifts two, so the two are neither the same recipe nor compared as one.

\textbf{Stage 4: Post-Hoc Composition. } With two policies never trained together, the agent swept a gradient-free merge over five scales on a short pilot; the shipped scale was chosen by us rather than by a gate or a gradient, and what that choice cost is dissected in Sec.~\ref{sec:traps}.

\subsection{Autonomy Traps: Outside the Contract's Reach}
\label{sec:traps}\label{sec:proxy_exploitation}
Beyond this progress, Table~\ref{tab:master_trajectory_summary}'s Failure Mode column reveals the traps of autonomy outside the contract's reach. Read against the design space of Sec.~\ref{sec:preliminaries}, each trap sits on one axis: data, proxy, and algorithm, and only the proxy trap is of the agent's own making.

\textbf{The Data Trap: Distribution leaks. } The measurement contract cannot guard a channel it never reads. Explicit clauses forbid test leakage, and every LM candidate passed through them. The leak came through the harness's own prompting channel instead: orchestrator guidance carried the held-out set verbatim into the proposer's context window, and no clause fired, because the contract audits candidates, not prompts. In a fully automated pipeline, boundary clauses fail without hard isolation of every channel the agent reads.

\textbf{The Proxy Trap: Reward hacking. } The agent optimizes the proxy score, not the audio. The same autonomy that, unprompted, replaced a speaker encoder circular with the generator it scored also read a probe's zero spread as noise and built a threshold-shaping term on an ungated proxy, forcing the band mean above the base in tens of steps. Table~\ref{tab:contrast} scores the same audio twice: on Mandarin, where the shaped reward was read, the proxy overstates the held-out gain by up to 56\%; on English, which the shaped term never read, the two instruments agree, and the unshaped speaker term does not inflate.

\textbf{The Algorithm Trap: Composition failures. } The literature assumes that LM RL and decoder RL improve different things and thus add. The agent tuned both, so we could test it. Merging its two policies shifted the upstream distribution: the decoder had been optimized on the base LM's tokens, and the gradient-free merge pushed speaker similarity below the base and gave back part of the LM policy's intelligibility gain (Table~\ref{tab:master_trajectory_summary}). The merge scale was chosen on a short pilot where the proxy kept rising past the held-out peak; the one decision without a gradient was ours, and it failed.

\subsection{Workspace Controllability: What Held, and What Slipped}
\label{sec:control}
Autonomous optimization (Sec.~\ref{sec:overall_performance}, \ref{sec:autonomy}) exposed ungated proxies and unread channels (Sec.~\ref{sec:traps}); we now audit the shared workspace (Sec.~\ref{sec:method}) for which controls held firm and which slipped. 

\textbf{Measurement Contract. } 
The preventative rules yielded mixed results. The Constraints and the LM gate functioned as intended, rejecting every failed LM candidate (e.g., $i_{16}$--$i_{19}$; Table~\ref{tab:master_trajectory_summary}), and repository-root access was never abused. However, the Proxies slipped at the pivot: the contract's scope never covered flow candidates, permitting reward hacking on ungated in-loop signals. Similarly, the Observers' sealed set was compromised by channel blindness: the held-out set leaked through the prompting channel, and the sealed-test clause ran 26 times but caught nothing. 

\textbf{Trajectory Memory. } Where prevention failed, the runtime log succeeded. The memory's append-only record of every recipe, score, and trace facilitated post-hoc auditability: it allowed us to trace and recover the nine tainted records the contract had inadvertently passed. 

\section{Conclusion}
We draw four findings in response to the question ``Can LLM agents automate reinforcement learning for text-to-speech?'' First, the pipeline automates: the agent recovers an underspecified method by measurement, improves it, and, given a symptom and no method, changes carrier unprompted, cutting Bad cases from $13.4\%$ to $6.1\%$. Second, a data trap: the held-out set leaked through the one channel the contract never read. Third, a proxy trap: the binding constraint is measurement rather than reasoning: the agent's shaping term inflated its proxy by up to $56\%$ where it was scored. Finally, an algorithm trap: gains do not add up: merging the policies caused a distribution shift that cost intelligibility against the LM policy and speaker similarity against the base. Taken together, these findings indicate that an agent automates RL post-training only as far as its harness measures; the harness, unlike the model, is ours to fix.

\section{Acknowledgment}
Generative AI was used solely for paper clarity. 
By exposing the strengths and weaknesses of LLM agents, we aim to help readers leverage them for accelerating code development and research ideation. 
The authors thoroughly verified all agent-generated results present in this paper and assume full responsibility. 

\bibliographystyle{IEEEbib}
\bibliography{strings,paper}

\end{document}

%% file: Figure/figure1new.tex
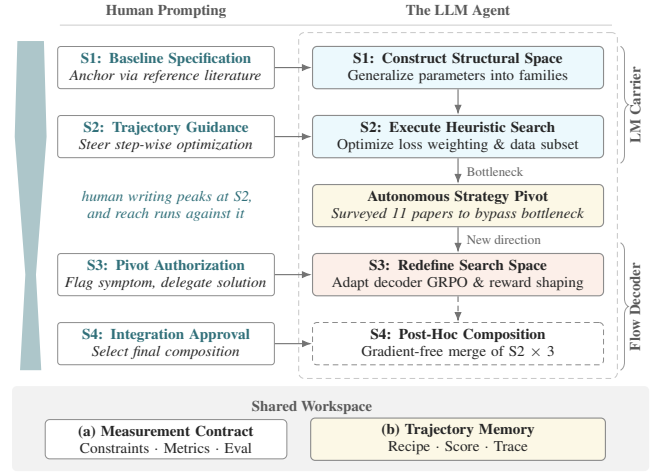
\begin{figure}[t]
\centering
\resizebox{0.98\columnwidth}{!}{% 
\begin{tikzpicture}[
  x=1mm, y=1mm, 
  font=\footnotesize, 
  every node/.append style={text=black!85},
  hum/.style={draw=black!45, rounded corners=1.6pt, fill=white, align=center,
              inner xsep=4pt, inner ysep=3.2pt, text width=38mm}, 
  stg/.style={draw=black!45, rounded corners=2.4pt, align=center,
              inner xsep=4pt, inner ysep=3.2pt, text width=52mm}, 
  lmn/.style={stg, fill=palSky!20, draw=black!45, semithick},
  fln/.style={stg, fill=palPeach!22, draw=black!45, semithick},
  mrg/.style={stg, fill=white, draw=black!45, densely dashed, text width=52mm}, 
  band/.style={rounded corners=2.4pt, align=center,
               inner xsep=5pt, inner ysep=3.2pt, font=\footnotesize}, 
  ar/.style={-{Latex[length=1.9mm,width=1.4mm]}, semithick, black!55},
  lbl/.style={font=\footnotesize, inner sep=1pt}, 
  hdr/.style={font=\footnotesize\bfseries, text=black!65}, 
  brc/.style={decorate, decoration={brace, amplitude=1.5mm, raise=0.6mm}, black!40}
]

% ---------- 標題與分隔線 (Y 座標從 72/69.5 往上推至 74.5/72.0 騰出空間) ----------
\node[hdr] at (22,74.5) {Human Prompting};
\node[hdr] at (77,74.5) {The LLM Agent}; 
\draw[black!20] (2,72.0) -- (106,72.0); 

% ---------- human writing background ----------
\fill[palTeal!40] (-6,69) -- (-6.5,51) -- (-4.5,25) -- (-5.5,7) --
                  (-1.5,7) -- (-2.5,25) -- (-0.5,51) -- (-1.0,69) -- cycle;

% ---------- the human branch ----------
\node[lbl, align=center, text=palTeal, text width=34mm] at (22,38)
     {\textit{human writing peaks at S2,}\\\textit{and reach runs against it}};
     
\node[hum] (h1) at (22,64) {\textcolor{palTeal}{\bfseries S1: Baseline Specification}\\\textit{Anchor via reference literature}};
\node[hum] (h2) at (22,51) {\textcolor{palTeal}{\bfseries S2: Trajectory Guidance}\\\textit{Steer step-wise optimization}};
\node[hum] (h3) at (22,25) {\textcolor{palTeal}{\bfseries S3: Pivot Authorization}\\\textit{Flag symptom, delegate solution}};
\node[hum] (h4) at (22,12) {\textcolor{palTeal}{\bfseries S4: Integration Approval}\\\textit{Select final composition}};

% ---------- the agent branch ----------
\node[lmn] (q1) at (77,64) {\textbf{S1: Construct Structural Space}\\Generalize parameters into families};
\node[lmn] (q2) at (77,51) {\textbf{S2: Execute Heuristic Search}\\Optimize loss weighting \& data subset};

\node[draw=black!45, semithick, fill=palYellow!25, rounded corners=2.4pt, inner xsep=4pt,
      inner ysep=3.2pt, align=center, text width=52mm] (piv) at (77,38) 
     {\textbf{Autonomous Strategy Pivot}\\\textit{Surveyed 11 papers to bypass bottleneck}};
     
\node[fln] (q3) at (77,25) {\textbf{S3: Redefine Search Space}\\Adapt decoder GRPO \& reward shaping};
\node[mrg] (mg) at (77,12) {\textbf{S4: Post-Hoc Composition}\\Gradient-free merge of S2 $\times$ 3};

% ---------- arrows ----------
\draw[ar] (h1) -- (q1);
\draw[ar] (h2) -- (q2);
\draw[ar] (h3) -- (q3);
\draw[ar] (h4) -- (mg);
\draw[ar] (q1) -- (q2);
\draw[ar] (q2) -- (piv) node[lbl, pos=0.5, right, xshift=1.2mm, fill=white, inner sep=1pt, text=black!60] {\scriptsize Bottleneck};
\draw[ar] (piv) -- (q3) node[lbl, pos=0.5, right, xshift=1.2mm, fill=white, inner sep=1pt, text=black!60] {\scriptsize New direction};
\draw[ar, densely dashed] (q3) -- (mg);

% ---------- background panel (清理了重複的 inner sep，統一設為 2.5mm) ----------
\begin{scope}[on background layer]
\node[draw=black!22, densely dashed, rounded corners=3pt, fit=(q1)(mg), inner sep=2.5mm] (panel) {};
\end{scope}

% ---------- right side labels ----------
\draw[brc] (107,68) -- (107,46);
\node[lbl, rotate=90, text=black!65] at (110.4,57) {\textbf{LM Carrier}};
\draw[brc] (107,31) -- (107,7);
\node[lbl, rotate=90, text=black!65] at (110.4,19) {\textbf{Flow Decoder}};

% ---------- the environment ----------
\begin{scope}[on background layer]
\fill[black!5, rounded corners=2.5pt] (-7,4.0) rectangle (113,-12.0); 
\end{scope}

\node[hdr, font=\footnotesize\bfseries] at (49.5, 0) {Shared Workspace}; 

\node[band, fill=white, draw=black!45, text width=42mm] (ct) at (22,-6.0)
    {\textbf{(a) Measurement Contract}\\ Constraints $\cdot$ Metrics $\cdot$ Eval}; 

\node[band, fill=palYellow!25, draw=black!45, semithick, text width=52mm] (lg) at (77,-6.0) 
     {\textbf{(b) Trajectory Memory}\\ Recipe $\cdot$ Score $\cdot$ Trace};

\end{tikzpicture}}
\vspace{-0.5em}
\caption{The human-agent collaborative AgenticTTS-Forge workflow.}
\label{fig:system}
\end{figure}

%% file: Table/table_roles.tex
\begin{table}[t]
\fontsize{7.5}{8.6}\selectfont
\setlength{\tabcolsep}{3pt}
\renewcommand{\arraystretch}{1.25}
\centering
\renewcommand\tabularxcolumn[1]{>{\arraybackslash}m{#1}}
\begin{threeparttable}
\caption{
Measurement contract: in-loop proxies and held-out observers. (LM Carrier introduced in S1–S2; Flow Carrier in S3). 
% Instantiation of the measurement contract: in-loop proxies and held-out observers per quantity 
% (flow column introduced by the agent at Stage 3; held-out scores in Table~\ref{tab:main}).
}
\label{tab:roles}
\begin{tabularx}{\columnwidth}{@{} l >{\hsize=0.92\hsize\raggedright\arraybackslash}X >{\hsize=0.92\hsize\raggedright\arraybackslash}X >{\hsize=1.16\hsize\raggedright\arraybackslash}X @{}}
\toprule
\multirow{2}{*}{Quantity} & \multicolumn{2}{c}{Proxies (In-loop)} & \multicolumn{1}{c}{\multirow{2}{*}{\begin{tabular}[c]{@{}c@{}}Observers\\(Held-out)\end{tabular}}} \\
\cmidrule(lr){2-3}
 & LM pairs / gate & Flow reward & \\
\midrule
Intellig. & pairs: SenseVoice\newline ASR~\cite{funaudiollm2024}; gate:\newline SenseVoice CER & SeACo-Paraformer\newline zh CER~\cite{shi2024seaco};\newline en unscored & Whisper~\cite{radford2023whisper} (en);\newline Paraformer-zh\newline \cite{gao2022paraformer,gao2023funasr} (zh) \\
\arrayrulecolor{black!35}\specialrule{0.3pt}{1pt}{1pt}\arrayrulecolor{black}
SECS & gate: CAM++~\cite{wang2023campplus} & ERes2Net~\cite{chen2023eres2net} & WavLM-large~\cite{chen2022wavlm} \\
\arrayrulecolor{black!35}\specialrule{0.3pt}{1pt}{1pt}\arrayrulecolor{black}
MOS & gate: UTMOS22~\cite{saeki2022utmos} & threshold-shaped\newline UTMOS22 & UTMOSv2~\cite{baba2024utmosv2} \\
\arrayrulecolor{black!35}\specialrule{0.3pt}{1pt}{1pt}\arrayrulecolor{black}
Bad case & not gated & not in reward & MOS $<$ 3 or WER/CER\newline $>$ 3\% (6\% on hard) \\
\bottomrule
\end{tabularx}
\begin{tablenotes}[flushleft]\scriptsize
\item Proxies either carry a gradient signal or only gate LM candidates (512-condition gate, one regression cap each); a gate has no gradient but is not held out either.
% \item[$\dagger$] Shares checkpoint lineage with the flow reward's ASR.
\item The LM gate also read parselmouth's F0 correlation (FPC)~\cite{zhang2024amphion}, not reported.
\end{tablenotes}
\end{threeparttable}
\end{table}

%% file: Table/table_main.tex
\begin{table*}[t]
\renewcommand{\arraystretch}{1.08}
\fontsize{8.0}{9.2}\selectfont
\caption{Objective evaluation on Seed-TTS-Eval; lower block scored by the held-out observers of Table~\ref{tab:roles}. Absolute scores are comparable only within each block (different inference conditions and MOS scorers). Teal marks gains over the released CosyVoice2 as run by us (Our Base).}
\vspace{0.2em}
\label{tab:main}
\begin{tabularx}{\textwidth}{@{} l *{10}{>{\centering\arraybackslash}X}}
\toprule
\multirow{2}{*}{System} & \multicolumn{3}{c}{Intelligibility} & \multicolumn{3}{c}{SECS} & \multicolumn{3}{c}{MOS} & \multirow{2}{*}{\begin{tabular}[c]{@{}c@{}}Bad\\case (\%)\end{tabular}} \\
\cmidrule(lr){2-4}\cmidrule(lr){5-7}\cmidrule(lr){8-10}
 & en~WER & zh~CER & hard~zh & en & zh & hard~zh & en & zh & hard~zh & \\
\midrule
\rowcolor{black!8}
\multicolumn{11}{c}{\textit{Expert-Tuned Preference Optimization on CosyVoice2 (Reported by Yao et al.~\cite{fpo2025})}} \\
{\scriptsize\itshape Reported models} & {\scriptsize\itshape Whisper} & \multicolumn{2}{c}{\scriptsize\itshape Paraformer-zh} & \multicolumn{3}{c}{\scriptsize\itshape WavLM-large} & \multicolumn{3}{c}{\scriptsize\itshape UTMOS22, all sets pooled} & {\scriptsize\itshape FPO def.} \\
CosyVoice2~\cite{du2024cosyvoice2} & 2.57 & 1.45 & 8.08 & 0.652 & 0.748 & 0.730 & \multicolumn{3}{c}{3.49 {\scriptsize(pooled)}} & 14.0 \\
\quad + SDPO~\cite{rafailov2023dpo} & 2.71 & 1.43 & 8.32 & 0.673 & 0.745 & 0.728 & \multicolumn{3}{c}{3.59 {\scriptsize(pooled)}} & 15.0 \\
\quad + UNO~\cite{chen2024uno} & 2.43 & 1.37 & 7.78 & 0.669 & 0.759 & 0.736 & \multicolumn{3}{c}{3.61 {\scriptsize(pooled)}} & 12.0 \\
\quad + RIO~\cite{hu2024rio} & 2.47 & 1.40 & 7.94 & 0.675 & 0.751 & 0.737 & \multicolumn{3}{c}{3.73 {\scriptsize(pooled)}} & 9.0 \\
\quad + Mask-DPO~\cite{gu2025maskdpo} & 2.51 & 1.42 & 7.92 & 0.671 & 0.754 & 0.733 & \multicolumn{3}{c}{3.64 {\scriptsize(pooled)}} & 13.0 \\
\quad + FPO~\cite{fpo2025} & 2.24 & 1.32 & 6.92 & 0.654 & 0.747 & 0.731 & \multicolumn{3}{c}{3.56 {\scriptsize(pooled)}} & 8.0 \\
\midrule
\rowcolor{palSky!22}
\multicolumn{11}{c}{\textit{Agent-Automated TTS Policies (Evaluated by Held-out Observers)}} \\
{\scriptsize\itshape Held-out observers} & {\scriptsize\itshape Whisper} & \multicolumn{2}{c}{\scriptsize\itshape Paraformer-zh} & \multicolumn{3}{c}{\scriptsize\itshape WavLM-large} & \multicolumn{3}{c}{\scriptsize\itshape UTMOSv2} & {\scriptsize\itshape FPO def.} \\
CosyVoice2 (Our Base) & 2.85 & 1.41 & 7.47 & 0.661 & 0.759 &0.737& 3.273 & 3.121 & 3.083 & 13.4 \\
\quad Stage 1: FPO Repro. ($i_{10}$) & \cellcolor{palTeal!18}2.79 & \cellcolor{palTeal!18}1.35 & \cellcolor{palTeal!18}6.90 & \cellcolor{palTeal!18}0.662 & 0.758 & \cellcolor{palTeal!18}0.738 & 3.272 & \cellcolor{palTeal!18}3.129 & \cellcolor{palTeal!18}3.097 & \cellcolor{palTeal!18}13.1 \\
\quad Stage 2: LM Carrier Policy ($i_{20}$) & \cellcolor{palTeal!18}2.47 & \cellcolor{palTeal!18}1.07 & \cellcolor{palTeal!18}6.41 & 0.658 & 0.755 & 0.735 & \cellcolor{palTeal!18}3.304 & \cellcolor{palTeal!18}3.146 & \cellcolor{palTeal!18}3.130 & 13.6 \\
\quad Stage 3: Flow Carrier Policy ($o_{23}$) & \cellcolor{palTeal!18}2.82 & \cellcolor{palTeal!18}1.37 & 7.48 & 0.660 & 0.754 & 0.727 & \cellcolor{palTeal!18}3.409 & \cellcolor{palTeal!18}3.304 & \cellcolor{palTeal!18}3.287 & \cellcolor{palTeal!18}6.1 \\
\quad Stage 4: Composition ($i_{20} \times o_{23}$) & \cellcolor{palTeal!18}2.49 & \cellcolor{palTeal!18}1.09 & \cellcolor{palTeal!18}6.51 & 0.656 & 0.749 & 0.724 & \cellcolor{palTeal!18}3.440 & \cellcolor{palTeal!18}3.325 & \cellcolor{palTeal!18}3.301 & \cellcolor{palTeal!18}5.4 \\
\bottomrule
\end{tabularx}
\end{table*}

%% file: Table/table1_condensed_v3.tex
% Condensed trajectory table, v3 (stacked key-value cells, stage shading).
% Usage in the main file:  \input{Table/table1_condensed_v3}
\begin{table*}[t!]
\centering
\fontsize{7.2}{8.2}\selectfont
\renewcommand{\arraystretch}{1.0}
\setlength{\tabcolsep}{3pt}
\caption{Overview of the four workflow stages (Fig.~\ref{fig:system}). Each stage traces the human prompting role, the published baseline recipe, the agent's autonomous edits across the design-space axes (Sec.~\ref{sec:preliminaries}), and the resulting empirical reach and failure modes.}
\label{tab:master_trajectory_summary}
\vspace{0.2em}
\begin{tabularx}{\textwidth}{@{}
  >{\hsize=0.84\hsize\raggedright\arraybackslash}X
  >{\hsize=0.92\hsize\raggedright\arraybackslash}X
  >{\hsize=1.30\hsize\raggedright\arraybackslash}X
  >{\hsize=1.02\hsize\raggedright\arraybackslash}X
  >{\hsize=0.92\hsize\raggedright\arraybackslash}X
  @{}}
\toprule
\textbf{Human Prompting} & \textbf{Published Recipe} & \textbf{Agent's Edit} & \textbf{Agent's Reach} & \textbf{Agent's Failure} \\
\midrule
\rowcolor{palSky!22}
\multicolumn{5}{c}{\textit{Stage 1, Construct Structural Space (LM carrier, $i_1$--$i_{10}$): 10 candidates, 25 attempts, 1 admitted, 33 GPU-h.}} \\
\textbf{Baseline Specification}\newline FPO paper only; observer mode\newline 5 manual unblocks; 1.4k chars/day &
\textbf{Disclosed (7):} 1{,}000 pairs, 100 speakers, gap 0.3, batch 64, 2 epochs, lr 2e-6, AdamW\newline \textbf{Unstated:} 14 &
\textbf{Data:} 462-pair pilot $\to$ 1{,}155 predicted $\to$ 967 (quality gate) $\to$ 923 (exact mask) $\to$ 920 trained\newline \textbf{Algo:} FPO form kept; 14 unstated $\to$ 21 search coordinates &
\textbf{Result:} hard zh CER 7.47$\to$6.90, zh CER 1.41$\to$1.35 (inside seed band)\newline \textbf{Recovered:} 45\% of FPO's zh CER gain, 5\% of its Bad case drop &
\textbf{None:} 5 halts = intentional pauses\newline \textbf{Cost:} 33 GPU-h for a gain inside seed variance \\
\specialrule{0.4pt}{0.6pt}{0.6pt}
\rowcolor{palSky!22}
\multicolumn{5}{c}{\textit{Stage 2, Execute Heuristic Search (LM carrier, $i_{11}$--$i_{31}$): 21 candidates (31 with Stage 1), 3 admitted; chain closes at $i_{20}$.}} \\
\textbf{Trajectory Guidance}\newline 1 falsifiable mutation/round\newline 1.8k chars/day (peak); harness 203k chars &
\textbf{Objective:} token-wise sum, no margin\newline \textbf{Data:} pairs generated once, never refreshed; starts at reference &
\textbf{Data:} $i_{20}$ 920 cached $\to$ 360 on-policy pairs (205 en, 155 zh)\newline \textbf{Algo:} $i_{14}$ equal-pair masked mean; $i_{15}$ margin $-\log\sigma(\beta(\Delta-m))$; warm vs.\ frozen reference &
\textbf{Result:} zh CER 1.41$\to$1.07, en WER 2.85$\to$2.47, hard zh 7.47$\to$6.41; tail 13.6\% unmoved\newline \textbf{Source:} data cut, not objective edits &
\textbf{Distribution Leak:} held-out set in proposer context ($i_{29}$--$i_{31}$)\newline \textbf{Rejected:} $i_{16}$--$i_{19}$ at the 512-condition gate \\
\specialrule{0.4pt}{0.6pt}{0.6pt}
\rowcolor{palPeach!22}
\multicolumn{5}{c}{\textit{Stage 3, Redefine Search Space (flow carrier, $g_1$--$o_{45}$): 45 candidates; 27 probes (depth $\le 5$), 18 full, 9 to 500 steps.}} \\
\textbf{Pivot Authorization}\newline symptom only; survey unrequested\newline 0.7k chars/day &
\textbf{FlowTTS-GRPO:} GRPO on decoder LoRA; CosyVoice3/F5-TTS\newline \textbf{Reward:} $z(\mathrm{SECS}) + z(1{-}\mathrm{CER}) + 0.4\,z(\mathrm{DNSMOS})$, linear &
\textbf{Data:} 192 conditions, UTMOS22 bands (dead $<3.0$)\newline \textbf{Proxy:} DNSMOS $\to$ UTMOS22 shaped $\sigma((u{-}3)/0.15)$, $w{=}0.4$; SeACo zh CER\newline \textbf{Algo:} decoder bottleneck $\to$ 11-paper survey $\to$ stochastic ODE sampler &
\textbf{Result:} 22/24 dead conditions revived at 500 steps; band mean UTMOS22 2.75$\to$3.44\newline \textbf{Held-out:} Bad case 13.4$\to$6.1\%; zh CER at base &
\textbf{Reward Hacking:} UTMOS22 overstates held-out UTMOSv2 gain by 43--56\% (zh only)\newline \textbf{Gap:} no flow candidate contract-gated \\
\specialrule{0.4pt}{0.6pt}{0.6pt}
\rowcolor{black!8}
\multicolumn{5}{c}{\textit{Stage 4, Post-Hoc Composition ($i_{20} \times o_{23}$, scale $s \in [0,1]$): 5 scales, 60-step pilot, no training.}} \\
\textbf{Integration Approval}\newline agent swept 5 scales; human picked one &
\textbf{None:} field assumes decoder RL + LM RL add up; never tested &
\textbf{Merge:} gradient-free $W + (\alpha/r)BA\,s$ over 352/1{,}121 decoder tensors\newline \textbf{Shipped:} $s{=}1.0$, past the held-out WavLM peak at 0.5 &
\textbf{Result:} Bad case 5.4\% (lowest row); UTMOSv2 above both parents, all 3 languages\newline \textbf{Sweep:} proxy rises at every $s$; WavLM peaks at 0.5 &
\textbf{Composition Failure:} zh CER $+0.016$ vs.\ $i_{20}$; WavLM $<$ base\newline \textbf{Cause:} LM change put decoder off-distribution \\
\bottomrule
\end{tabularx}
\end{table*}

%% file: Table/table_contrast_v2.tex
\begin{table}[!tb]
\renewcommand{\arraystretch}{1.08}
\fontsize{8.0}{9.2}\selectfont
\caption{Reward-hacking evaluation on in-loop proxies vs.\ held-out observers. Ratio denotes in-loop over held-out gain.}
\vspace{0.2em}
\label{tab:contrast}
\begin{tabularx}{\columnwidth}{@{} l l *{3}{>{\centering\arraybackslash}X} @{}}
\toprule
Loop role & Quantity & $\Delta$ in-loop & $\Delta$ held-out & Ratio \\
\midrule
\rowcolor{palPeach!22}
\multicolumn{5}{c}{\textit{Shaped reward (flow, $o_{23}$): gradient with threshold shaping}} \\
\quad zh & MOS & $+0.261$ & $+0.183$ & \textbf{1.43} \\
\quad hard zh & MOS & $+0.319$ & $+0.204$ & \textbf{1.56} \\
\addlinespace[2pt]
\rowcolor{palPeach!22}
\multicolumn{5}{c}{\textit{Unshaped reward (flow, $o_{23}$): gradient, no shaping}} \\
\quad en & SECS & $-0.006$ & $-0.001$ & -- \\
\quad zh & SECS & $-0.004$ & $-0.005$ & -- \\
\addlinespace[2pt]
\rowcolor{palPeach!22}
\multicolumn{5}{c}{\textit{Not in reward (flow, $o_{23}$): never read, no gradient}} \\
\quad en & MOS & $+0.121$ & $+0.136$ & 0.89 \\
\addlinespace[2pt]
\rowcolor{palSky!22}
\multicolumn{5}{c}{\textit{Gate only (LM, $i_{20}$): 0.03 cap, no gradient}} \\
\quad en & MOS & $+0.049$ & $+0.031$ & -- \\
\quad zh & MOS & $-0.004$ & $+0.025$ & -- \\
\quad hard zh & MOS & $-0.001$ & $+0.047$ & -- \\
\bottomrule
\end{tabularx}
\end{table}